\documentstyle[prd,aps,preprint]{revtex}

\begin{document}

\draft

\preprint{\font\fortssbx=cmssbx10 scaled \magstep2
\hbox to \hsize{
\special{
			      hscale=8000 vscale=8000
			       hoffset=-12 voffset=-2}
\hskip.5in \raise.1in\hbox{\fortssbx University of Wisconsin - Madison}
\hfill$\vbox{\hbox{\bf MADPH-95-886}
             \hbox{April 1995}}$ }
}

\title{Particle Production in Very High-Energy \\
Cosmic-Ray Emulsion Chamber Events: \\
Usual and Unusual Events}
\author{C. G. S. Costa and F. Halzen}
\address{Department of Physics, University of Wisconsin,
       Madison, WI 53706, USA}
\author{C. Salles}
\address{Department of Materials Science and Engineering, \\
University of Wisconsin, Madison, WI 53706, USA}
\date{\today}
\maketitle

\begin{abstract}
We show that a simple scaling model of very forward particle
production, consistent with accelerator and air shower data, can
describe all features of the very high-energy interactions recorded
with emulsion chambers. This is somewhat surprising after numerous
claims that the same data implied large scaling violations or new
dynamics. Interestingly, we cannot describe some of the Centauro
events, suggesting that these events are anomalous independently of
their well-advertised unusual features such as the absence of neutral
secondaries.
\end{abstract}

\section{Introduction}
In this paper we show that the simplest assumptions on particle
production in the very forward direction, Feynman scaling and
constant inelasticity, can provide a coherent description of air
shower and emulsion data. All features of our model are approximately
shared with QCD-inspired models\cite{halzen} of high-energy particle
interactions. The phase space populated by the secondaries in
very-high-energy cosmic-ray interactions is, unfortunately, not covered
by collider experiments which only view centrally produced particles.
We therefore start the discussion with the simplest ansatz for the
dynamics of particle production at large rapidities. In the end,
this guess will turn out to be adequate in describing a wide variety
of data.

\label{sec:theory}

In order to describe the underlying high-energy particle production
dynamics common to accelerator, air shower and emulsion chamber data,
we start with the simplest phenomenological model for the rapidity
density distribution of charged pion production which we parametrize
as
\begin{equation}
{dN\over dy} = x {dN\over dx} = a {(1-x)^n \over{x^m}} \;,
\label{rapdens}
\end{equation}
where $y$ is the rapidity of the produced secondaries. The Feynman
variable $x$ is given by the ratio of the secondary particle energy
$E$ to the incident energy $E_0$.

The $(1-x)$ power law in (\ref{rapdens}) describes the decrease in
the number of particles produced with large $x$ values, the $1/x$
term dictates the shape of the function at low $x$. Simple parton
counting rules~\cite{gunion} imply that $n\simeq$3 and $m=0-1/2$.
Eq.~(\ref{rapdens}) reflects the approximate Feynman scaling
behavior in the  fragmentation region (large $x,y$) expected
from QCD~\cite{halzen}.
Our basic assumptions imply that the inelasticity $K$, i.e. the
fraction of the collision energy going into the production of
secondaries, is independent of energy. QCD-inspired models predict
that $K=0.6-0.65$~\cite{gaisser,kopel,durand} at the energy region
of interest. We have to account for nuclear target effects which
can simply be done by allowing increased $m$-values to describe
the enhanced production of low-$x$ particles in nuclear collisions.
Also, nuclear effects increase the value of the
proton-air inelasticity~\cite{jaosh} to approximately $0.8$. This
simple picture is consistent with the little information one can
extract from accelerator results on the production of particles in
the very forward direction~\cite{UA7}. Especially in the high-energy
collider  experiments the phase space relevant to cosmic-ray
data is  usually obscured by the beam-pipe~\cite{bjorken}.

With this background we may investigate air shower and emulsion
chamber results.

\section{Forward Particle Production and Air Showers}

A QCD-based Monte Carlo simulation of ultra-high-energy cosmic-ray
hadron interactions, which incorporates large $p_t$ jet production,
has been used by L.K.~Ding et al.~\cite{ding} (hereafter referred to
as DKTY) to perform a detailed description of air shower observations.
The rapidity distribution of secondary particles has been
derived in this analysis~\cite{fh-japan}, e.g.\ for interaction
energies of $10^{16}$~eV ($\sqrt s \sim 4.3$~TeV) as illustrated in
Fig.~\ref{fig:rapdis}. Also shown in the figure is an adequate
description of the resulting distribution using Eq.~(\ref{rapdens})
with $a=0.12$, $n=2.6$ and $m=1$, parameters quite consistent with
our expectations. The analysis of reference~\cite{ding} also shows
that particle production in the forward direction exhibits
approximate Feynman scaling, consistent with our simple assumptions.

The main point of this paper is that this model, in its simplest
form, describes all the details of emulsion chamber experiments.
This was somewhat surprising to us given the many claims of new
dynamics and non-scaling behavior in the literature based on
the same information; see {\it e.g.} Ref.~\cite{ohsawa-92}.
Interestingly, we cannot describe some of the Centauro events.
The rapidity distribution of the secondaries in these events
pegs them as being anomalous, independently of their
well-advertised unusual features such as the absence of
neutral secondaries.

\section{Emulsion Chamber Data}

Before proceeding with the study of emulsion chamber data, we first
comment on a crucial aspect of the calculation: the mean inelasticity
which is calculated through energy conservation (using charged
secondaries only)
\begin{equation}
{2\over3}\left< K \right> E_0 = \int_0^1 E_0 \; x \left(dN\over
dx\right) dx \;.
\label{conserv}
\end{equation}
The unphysical singularity of Eq.~(\ref{rapdens}) on $x=0$ is removed
by the threshold effects in the measurement. The lowest energy
particles in the cascade will not make it to observation level.
We implement this by making the $x$-distribution constant below a
certain value $x_0$, so that the mean inelasticity is
\begin{equation}
\left< K \right> = {2\over3} \; a \; \left[ \int_{x_0}^1
{(1-x)^n\over x^m}  dx
+ {(1+x_0)^n \over x_{0}^{m-1} } \right] \;.   \label{mean}
\end{equation}

The existence of a lower limit on $x$, resulting from inevitable
energy-loss in the air cascade before it reaches the detector, is
well known~\cite{ohsawa-87}. Emulsion chambers observe the debris of air
showers initiated by cosmic rays one or more nucleonic
mean-free-paths (mfp) above the detector. From the study of
Ref.~\cite{ohsawa-87} one concludes that, because of the strong energy
dissipation in the cascade, $x_0$ must be around $1\times10^{-3}$ and
$3\times10^{-2}$ for mountain altitude measurements. We can obtain
the same result from a back-of-the-envelope calculation. Take as an
example the Ursa Maior event observed by the Brazil-Japan
Collaboration~\cite{chin}. Triangulation of the secondary charged
tracks predicts that the first interaction occurred 2 or 3~km above
the chamber, which represents approximately 2 mfp. At the first
interaction the nucleon with energy $E_0$ releases on average around
${1\over2}E_0$ to particle production and therefore ${1\over6}E_0$ to
each pionic component ($\pi^0,\, \pi^\pm$). After 2 more
interactions, a charged pion would have on average $(E_0/6)^3$ so
that $x_0\sim (1/6)^3 \sim 5\times10^{-3}$. This value is consistent
with the one obtained in Ref.~\cite{ohsawa-87} and, if used in
Eq.~(\ref{mean}), leads to $\left< K \right> = 0.81$, matching the
expected value; see Section~\ref{sec:theory}.

We will investigate experimental data obtained in large emulsion
chambers, using rigorous analytical solutions of the diffusion
equations for the hadronic cascade induced by one single nucleon in
the atmosphere. The one-dimensional solution $F_H(E,E_0,t,t_0)$
enables us to calculate the integral energy spectrum of
hadrons~\cite{bell-1}
\begin{equation}
I_H(>E,E_0,t_f,t_0) = \int_E^{E_0} dE' F_H(E',E_0,t_f,t_0) \;,
\label{spectrum}
\end{equation}
which gives the number of hadrons with energy above $E$, at the
detection level $t_f$ for a cascade initiated by a nucleon with
energy $E_0$ at atmospheric depth $t_0$. The three-dimensional
solution $F_H(E,E_0,{\bf r},t,t_0)$ is used to calculate the
energy-weighted lateral spread~\cite{bell-2,bell-3}
\begin{eqnarray}
N_H(>Er,E_0,t_f,t_0) = && \int\int dE' d{\bf r'} \, H(E'r'-Er)
\nonumber \\ && \times \,
F_H(E',E_0,{\bf r'},t_f,t_0) \;, \label{spread}
\end{eqnarray}
which evaluates the number of hadrons with energy times radial
length greater than $Er$. The function $H(E'r'-Er)$ is the
Heaviside function.

The only free parameters are the primary particle energy $E_0$
and the atmospheric depth $t_0$ of the first interaction.
In our calculations we define the visible energy
$E_0^{(\gamma)} = k_\gamma E_0$
(with  $k_\gamma \sim \left< k_\gamma \right> = 0.25$)
and the atmospheric depths $t$ in units of nucleonic mfp, with $\lambda_N =
80\rm~g/cm^2$.
We use for the transverse momentum
$\left< p_t \right> = 0.4\rm~GeV/c$.

In Fig.~\ref{fig:ursa} we compare our results for the integral
spectrum of the event Ursa Maior~\cite{chin} detected at
Mt.~Chacaltaya (540~g/cm$^2$ or $t_f=6.75$). We obtain
$E_0^{(\gamma)} = 2930$~TeV, which corresponds in the center of mass
system (c.m.s.) to $\sqrt s = 4.7$~TeV. The determined depth of the
first interaction ($t_0=4.95$) is consistent with the
expectation that the event started about 2 mfp above the
chamber (see  Table~\ref{tab:emuls}).
There is no data available on the energy-weighted lateral spread
for this event.

In Fig.~\ref{fig:P3C7} we show (a)~the integral spectrum and (b)~the
energy-weighted lateral spread for the superfamily P3$'$-C5-505,
detected by the Pamir Collaboration~\cite{pamir} at the detection
level $t_f=7.45$ (or 596~g/cm$^2$).
In  the same figure we present the results for the event
Centauro\,VII, detected by the Brazil-Japan
Collaboration~\cite{brazjap}. For clarity, we have shifted in the
figure the data of Centauro\,VII by a factor 10.
The characteristics of both events are again summarized in
Table~\ref{tab:emuls}.

The investigation of Centauro\,I, also detected by the Brazil-Japan
Collaboration~\cite{lattes,navia}, is presented in
Fig.~\ref{fig:centauro1} (see also Table~\ref{tab:emuls}).
We observe that neither the integral spectrum,
Fig.~\ref{fig:centauro1}(a), nor the energy-weighted
lateral spread, Fig.~\ref{fig:centauro1}(b), are reproduced
by the analytical calculation.
We verified that the origin of this discrepancy cannot be
associated with an increase of the transverse momentum of the
secondaries, for which there is some experimental evidence.

\section{Conclusions}

Guided by QCD-inspired models, we proposed a simple
phenomenological model for the production of secondaries,
based on approximate scaling in the fragmentation region. The model
parametrization is consistent with the behaviour of air showers
at ultra-high energies. We subsequently applied the model to the
description of experimental data obtained in emulsion chambers at
different mountain altitudes. Both integral spectrum and lateral
spread of cosmic-ray superfamilies are successfully reproduced.

We noticed that Centauro\,VII is well described by the model.
 From our point of view, it is not really different from any other
event resulting from normal strong interactions. Actually,
it has been pointed out previously~\cite{bell-2} that this
event is not a ``hard'' Centauro, in the sense that it possesses
a comparable number of hadrons and gamma-rays arriving at the
top of the upper detection chamber.  On the other hand, the
investigation demonstrates that an event like Centauro\,I cannot
be described with the same simple hypotheses, suggesting
that this event is clearly anomalous, even before one
considers their charge-to-neutral ratio.

\acknowledgements

This research was supported in part by the
Brazilian agency CNPq, in part by the
U.S.~Department of Energy under Contract
No.~DE-AC02-76ER00881, by the National Science
Foundation under Contract No.~DMR-9319421 and the
University of Wisconsin Research Committee with funds
granted by the Wisconsin Alumni Research Foundation.

\newpage
\begin{table}[h]
\caption{\label{tab:emuls} Determined parameters of emulsion
chamber events analyzed.}
\smallskip
\centering
\begin{tabular}{lccc}
Event& $E_0^{(\gamma)}$ (TeV)& $t_0$ ($\lambda_N$ units)& $t_f-t_0$
($\lambda_N$ units)\\
\tableline
Ursa Maior& 2930& 4.95& 1.80\\
P3$'$-C5-505& 3220& 1.45& 6.00\\
Centauro\,VII& 7380& 1.05& 5.70\\
Centauro\,I& 1890& 5.75& 1.00\\
\end{tabular}
\end{table}

\begin{figure} \label{fig:rapdis}
\caption{Rapidity distribution of forward secondaries as
derived from air showers with energy near $10^{16}$~eV,
calculated by DKTY ($*$) and Eq.~(\protect\ref{rapdens}), with
$a=0.12$, $n=2.6$  and $m=1$ (solid line).}
\end{figure}

\begin{figure} \label{fig:ursa}
\caption{Integral energy spectrum for the Ursa Maior event
($\diamondsuit$), detected at Mt.~Chacaltaya~\protect\cite{chin},
compared to the analytical calculation (solid line) using
the $x$-distribution of  Eq.~(\protect\ref{rapdens}). The fitted
parameters $E_0^{(\gamma)}$ and $t_0$  are given in
Table~\protect\ref{tab:emuls}.}
\end{figure}

\begin{figure} \label{fig:P3C7}
\caption{Description of the superfamily P3$'$-C5-505 ($\bigcirc$),
detected by the Pamir Collaboration~\protect\cite{pamir}, and of the event
Centauro\,VII ($\bigtriangleup$), detected by the Brazil-Japan
Collaboration~\protect\cite{brazjap}. For illustrative purposes, the data
of Centauro\,VII have been shifted by a factor 10. Solid lines are
the analytical calculations for: (a)~Integral energy spectrum,
(b)~energy-weighted lateral spread;
parameters in Table~\protect\ref{tab:emuls}.}
\end{figure}

\begin{figure} \label{fig:centauro1}
\caption{Investigation of the event Centauro\,I
($\diamondsuit$), detected at Mt.~Chacaltaya~\protect\cite{lattes,navia}:
(a)~Integral energy spectrum, (b)~energy-weighted lateral spread.
Solid lines are the analytical calculations;
parameters in Table~\protect\ref{tab:emuls}.}
\end{figure}

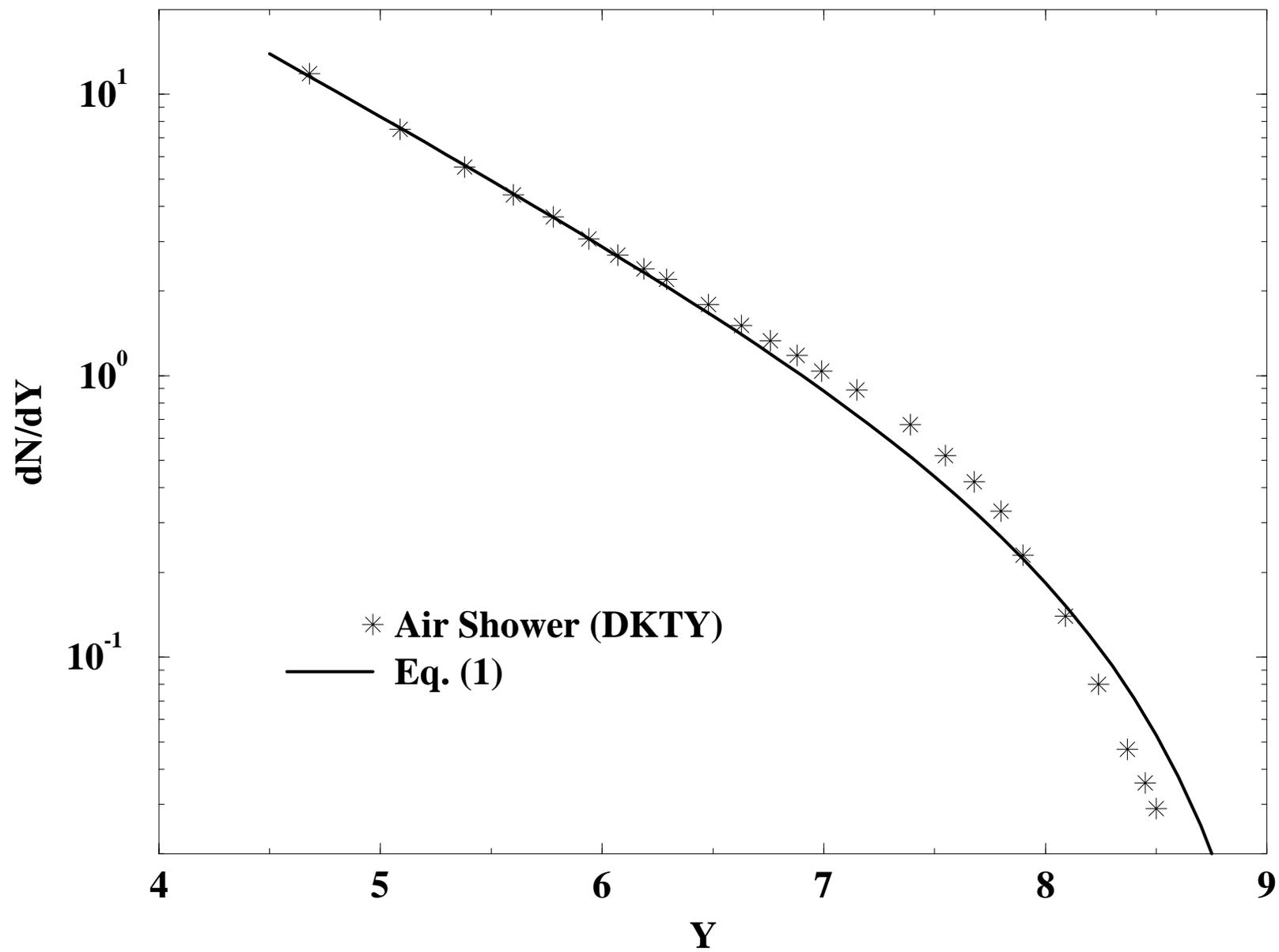
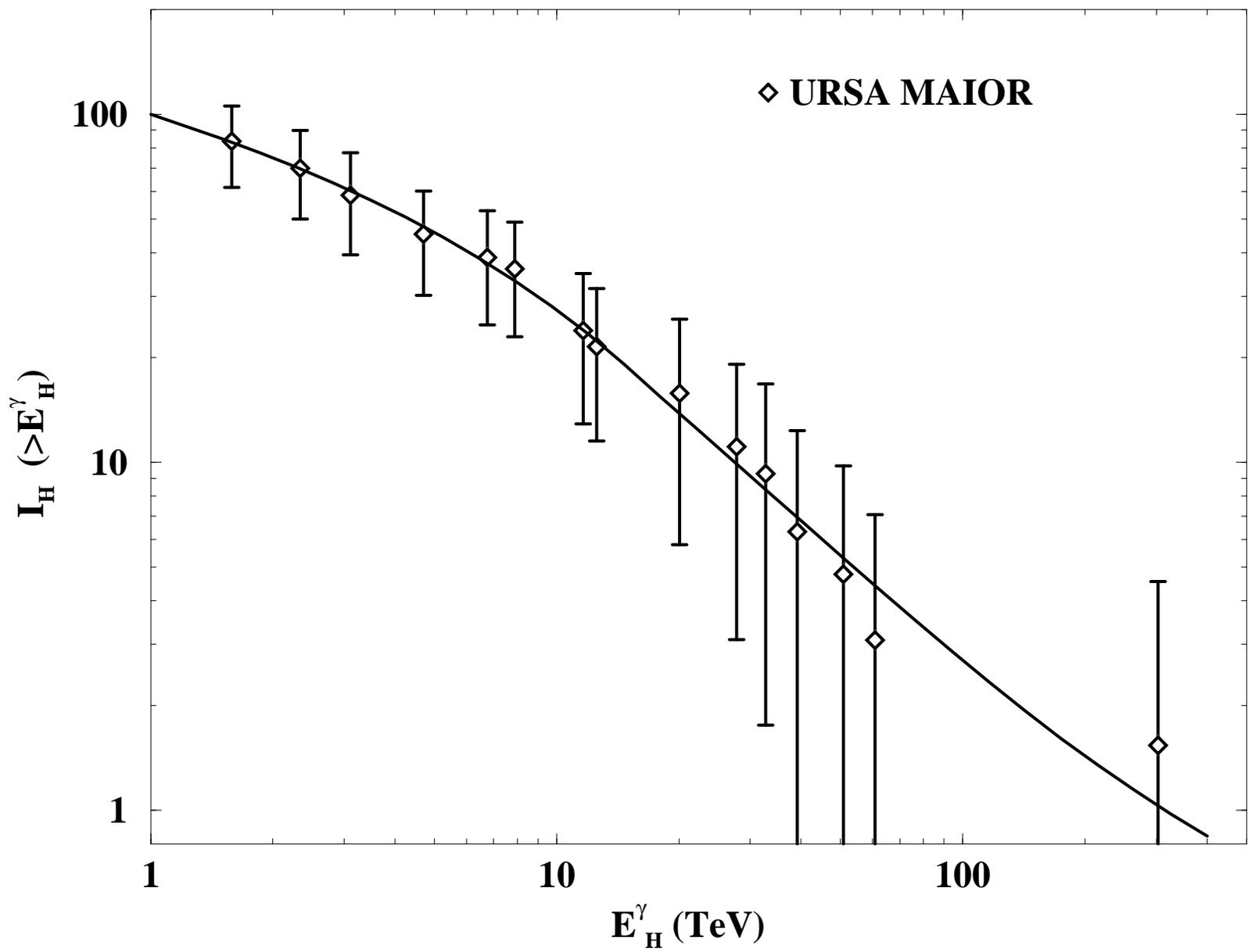
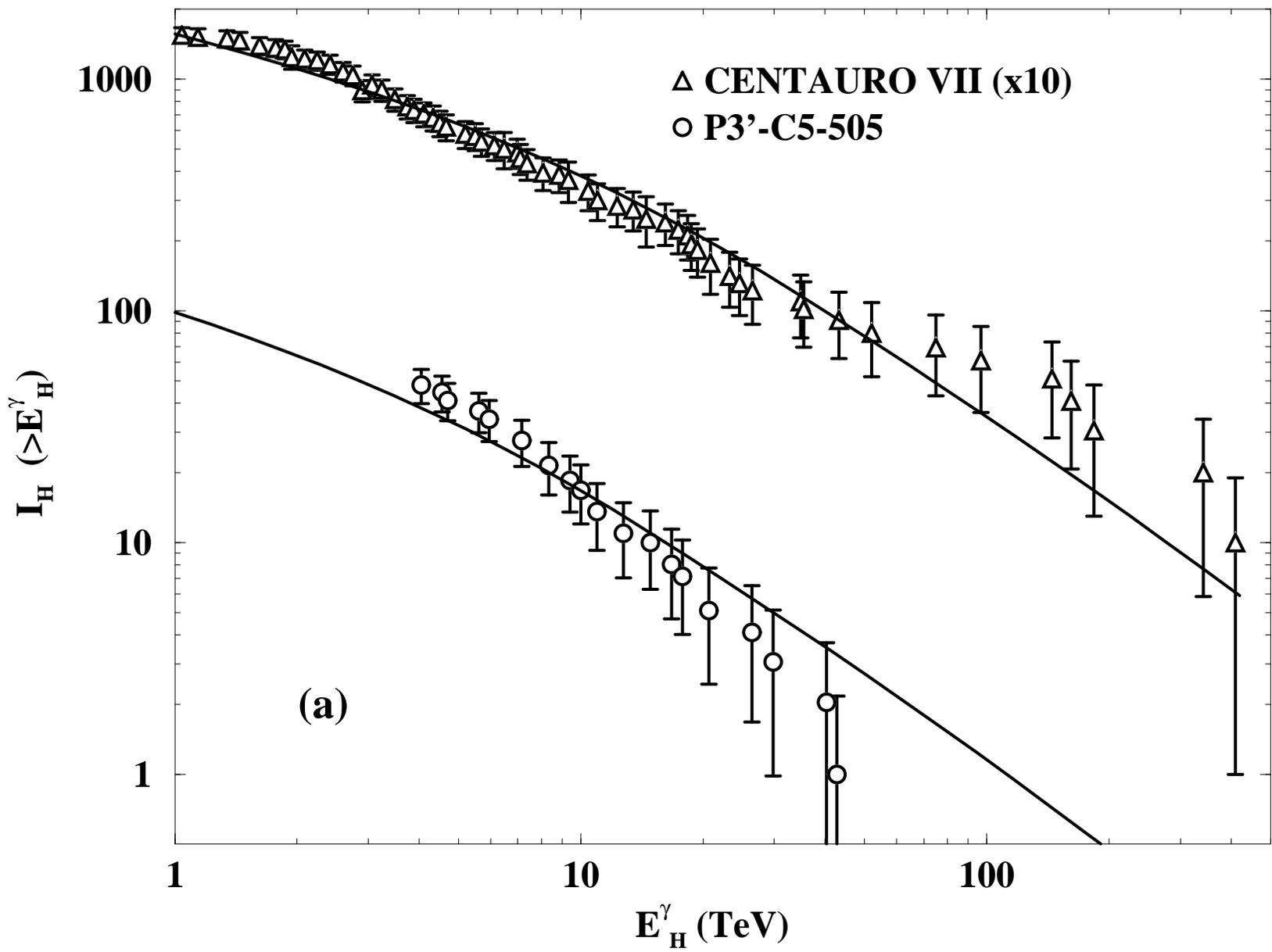
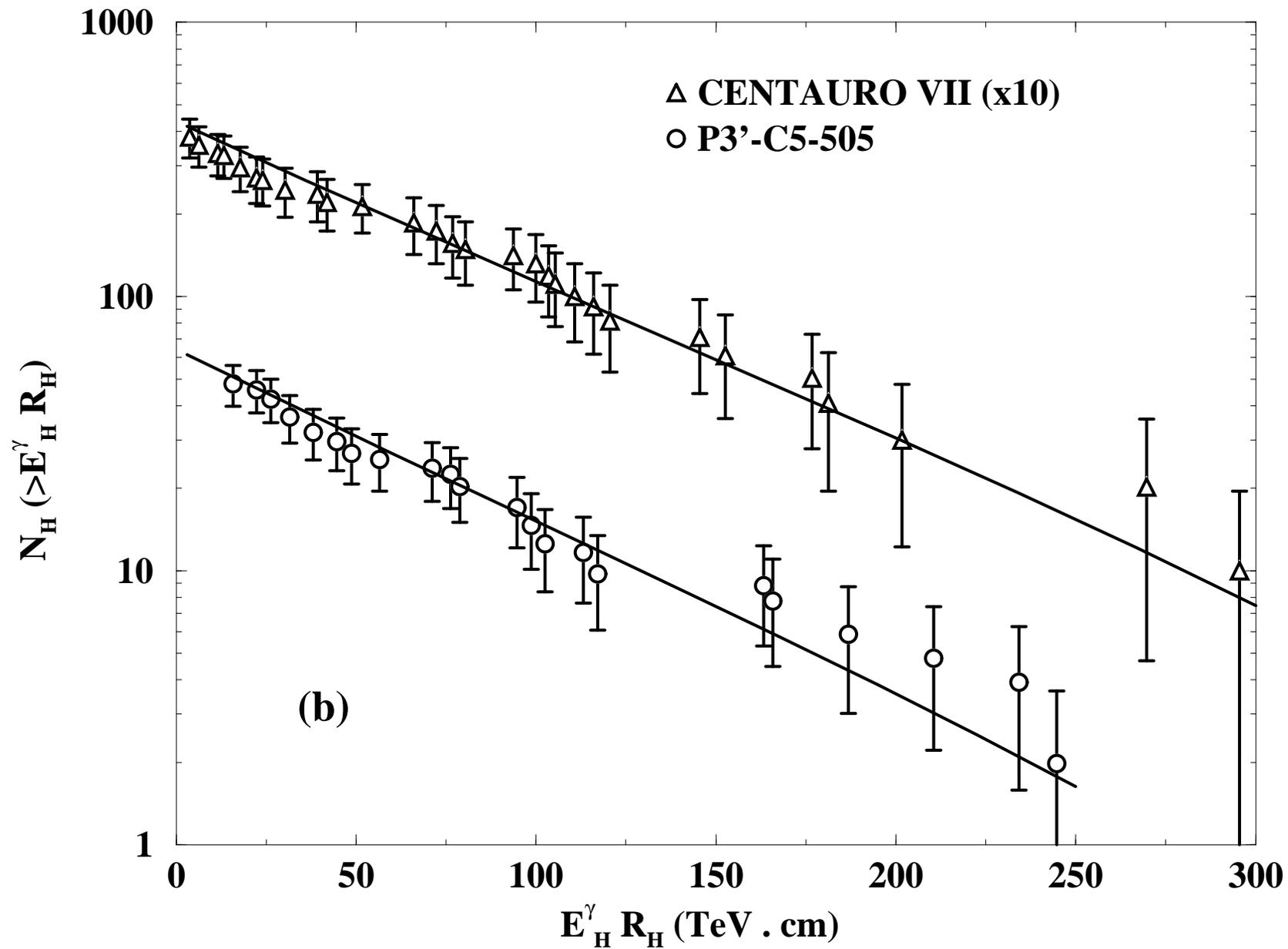
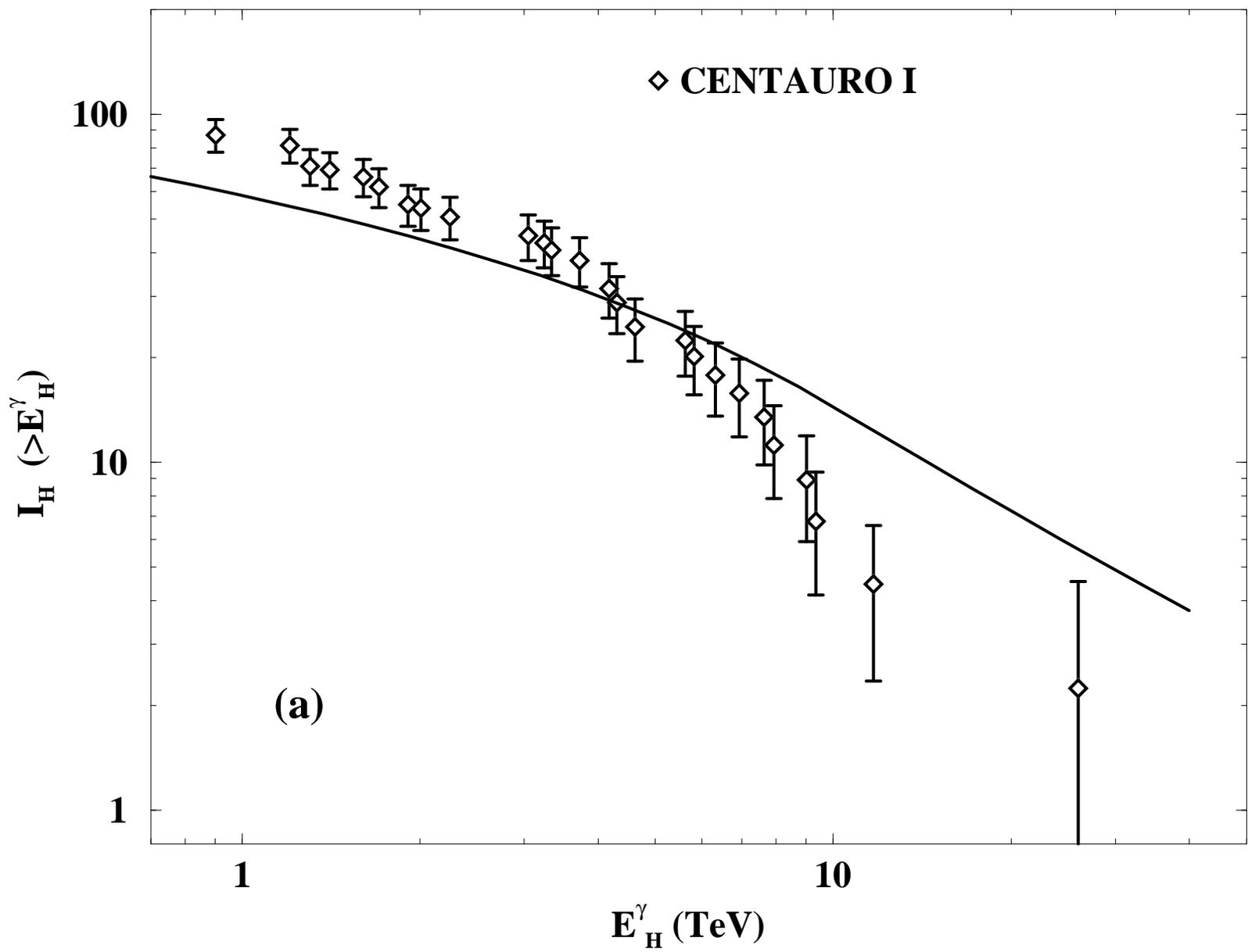
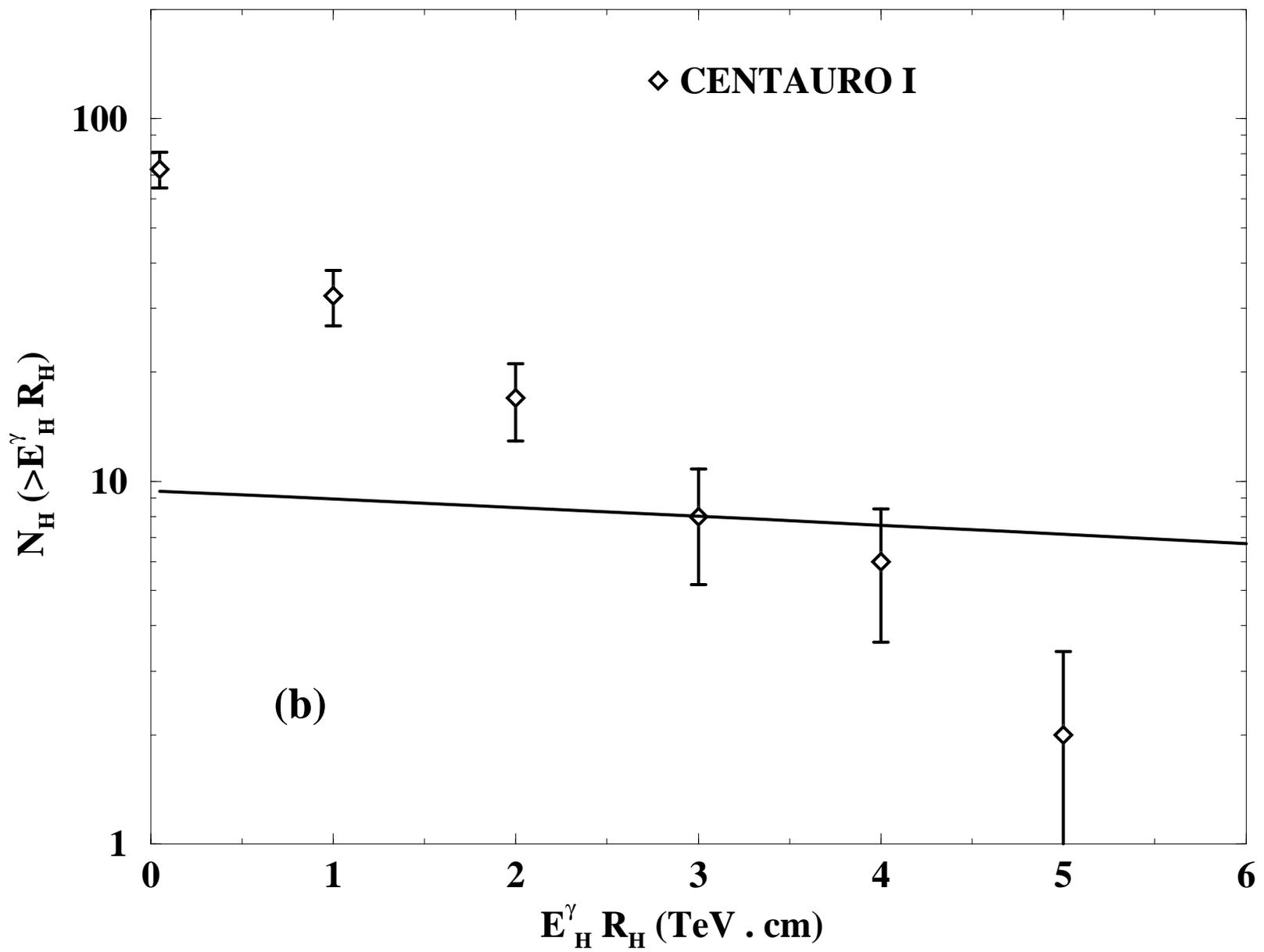
\end{document}